\documentclass[pra,unsortedaddress,superscriptaddress,a4paper,nofootinbib,%
nobalancelastpage,preprintnumbers,showpacs]{revtex4}%
\usepackage{amssymb,amsfonts,amsbsy}
\catcode`\@=11
\def\eqalign#1{\null\,\vcenter{\openup\jot\m@th
\ialign{\strut\hfil$\displaystyle{##}$&$\displaystyle{{}##}$\hfil
     \crcr#1\crcr}}\,}
\catcode`\@=12
\catcode`\@=11
\def\Eqalign#1{\null\,\vcenter{\openup\jot\m@th
\ialign{\strut\hfil$\displaystyle{##}$&$\displaystyle{{}##}$\hfil
&&\qquad\strut\hfil$\displaystyle{##}$&$\displaystyle{{}##}$\hfil
     \crcr#1\crcr}}\,}
\catcode`\@=12

  \usepackage[dvips]{graphicx}
                \DeclareGraphicsExtensions{.eps, .jpg}
\let\gs=\gtrsim
\def\vec#1{\boldsymbol #1}
\begin{document}
 \author{Monique Combescure}
 \affiliation{Institut de Physique Nucl\'aire de Lyon, CNRS-IN2P3 and Universit\'e Claude Bernard, 
 4, rue Enrico Fermi, 69622 Villeurbanne cedex, France}
 \author{Avinash Khare}
 \affiliation{Institute of Physics, Bubhaneswar, India}
 \author{Ashok Raina}
 \affiliation{Tata Institute of Fundamental Research,
School of Natural Sciences,
Homi Bhabha Road,
Mumbai 400005, India}
 \author{Jean-Marc Richard}
 \affiliation{LPSC--CNRS-IN2P3, 53, avenue des Martyrs, 38021 Grenoble cedex, France}
 \affiliation{Universit\'e Joseph-Fourier, Grenoble, France}
 \author{Carole Weydert}
  \affiliation{Universit\'e Joseph-Fourier, Grenoble, France}
 \title{Level rearrangement in exotic atoms and quantum dots}
 \begin{abstract}
A presentation and a generalisation are given of the phenomenon of level rearrangement, which occurs when an attractive long-range potential is supplemented by a short-range attractive potential of increasing strength. This problem has been discovered in condensate-matter physics and  has also been studied in the physics of exotic atoms. A similar phenomenon occurs in a situation inspired by quantum dots, where a short-range interaction is added to an harmonic confinement. 
\end{abstract}
 \maketitle  
 \section{Introduction}
 In 1959, Zel'dovich \cite{Zel59} discovered an interesting phenomenon while considering an excited electron in a semi-conductor. The model describing  the electron--hole system  consists of a Coulomb attraction modified at short-distance \cite{kolomeisky:022721}. A similar model  is  encountered in the physics of exotic atoms: if an electron is substituted by a negatively-charged hadron, this hadron feels both the Coulomb field and the strong interaction of the nucleus. The Zel'dovich effect has also been discussed for atoms in a strong magnetic field \cite{Karnakov}.
 
\begin{figure}[!hbct]
\begin{minipage}{.65\textwidth}
\centerline{\scalebox{.6}{\includegraphics{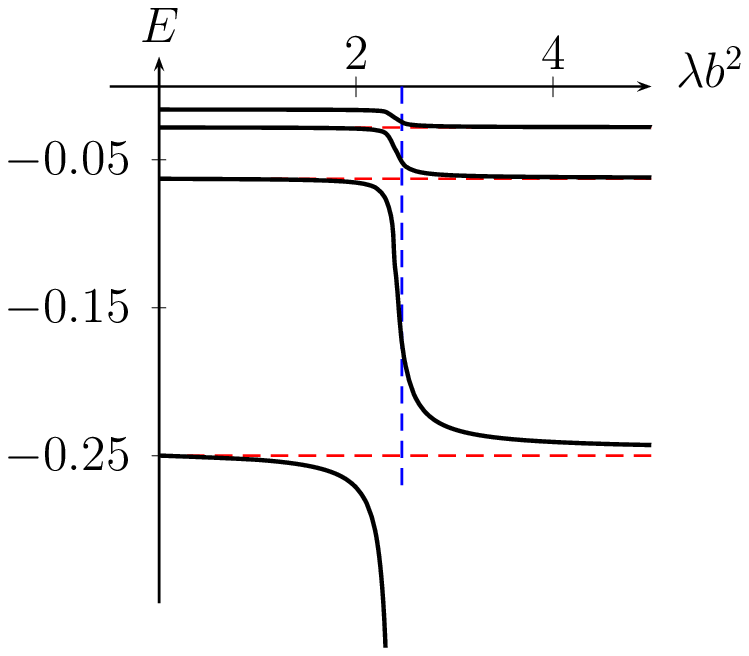}}
\hspace*{.2cm}
\scalebox{.6}{\includegraphics{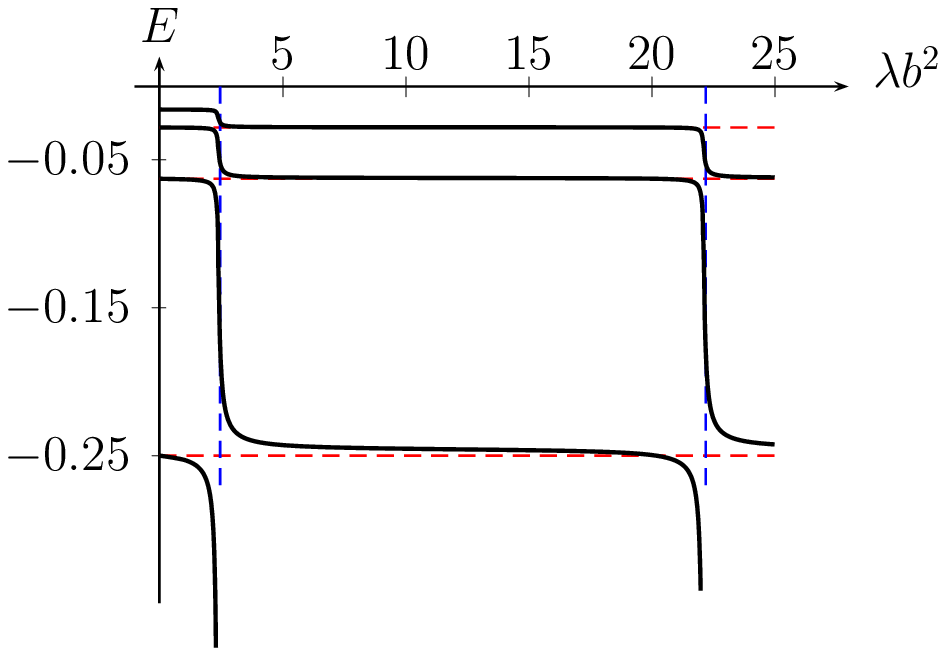}}}
\end{minipage}
\begin{minipage}{.34\textwidth}
\caption{\label{CoulSW} S-wave spectrum of the Coulomb potential (rescaled to $e^2=\hbar^2/(2\mu)=1$) modified by an attractive square well of radius $b=0.01$, and variable strength $\lambda$: first rearrangement (left) and second rearrangement (right). The dotted lines show the pure Coulomb energies and the coupling thresholds at which the square well alone supports one or two S-wave bound states.}
\end{minipage}
\end{figure}

 Zel'dovich  \cite{Zel59} and later Shapiro and his collaborators \cite{Kudryavtsev:1978af,Shapiro:1978wi} look at how the atomic spectrum evolves when the strength of the short-range interaction is increased, so that it becomes more and more attractive. The first surprise, when this problem is encountered, is that the atomic  spectrum is almost unchanged even so the nuclear potential at short distance is much larger than the Coulomb one. When the strength of the short-range interaction reaches a critical value, the ground state of the system leaves suddenly the domain of typical atomic energies, to become a nuclear state, with large negative energy.
 The second surprise is that, simultaneously, the first radial excitation leaves the range of values very close to the pure Coulomb 2S energy and drops towards (but slightly above) the 1S energy. In other words, the ``hole'' left by the 1S atomic level becoming a nuclear state is immediately filled by the rapid fall of the 2S. Similarly, the 3S state replaces the 2S, etc.
 This is why the process is named ``level rearrangement''. An illustration is given in Fig.~\ref{CoulSW}, for a simple square well potential supplementing a Coulomb potential.
 
 In this article, the phenomenon of level rearrangement is reviewed and generalised,  to account for cases where the narrow potential is located anywhere in a wide attractive well. An example is provided by a short-range pairwise interaction acting between two particles confined in an harmonic  potential, a problem inspired by the physics of quantum dots. The basic quantum mechanics of exotic atoms will be briefly summarised,  in particular with a discussion about the Deser--Trueman formula that gives the energy shift of exotic atoms in terms of the scattering length of the nuclear potential. A pedestrian derivation of this formula will be given in Appendix, which extents its validity beyond the case of exotic atoms. The link from the Coulomb to the harmonic cases will also be discussed in light of the famous 
 Kustaanheimo--Stiefel (KS) transformation, which is reviewed in several papers (see, e.g., \cite{Mavromatis:1998} and refs.\ there) and finds here an interesting application.
 
 The discussion is mainly devoted to one-dimensional problems or to S-states ($\ell=0$) in three dimensions. In Sec.~\ref{se:order}, it is extended to the first P-state (2P), and it is shown that the rearrangement is much sharper for P and higher $\ell$ states than for S states.

\section{Coulomb potential plus short-range attraction}\label{se:Coul-sr}
The simplest model of exotic atoms corresponds to the Hamiltonian
 \begin{equation}\label{basic-H}
 H=-\Delta -{1\over r} +\lambda v(r)~,
 \end{equation}
where $v(r)$ has a range that is very short as compared to the Bohr radius of the pure Coulomb problem.  Throughout this paper,  the energy units are set such that $\hbar^2/(2\mu)=1$, where $\mu$ is the reduced mass. In (\ref{basic-H}) the scaling properties of the Coulomb interaction are also used to fix the elementary charge $e=1$, without loss of generality. The study will be restricted here to S-wave states.  The case of P-states or higher waves  is briefly discussed in Sec.~\ref{se:order}.

As an example, a simple square well $v(r)=-\theta(b-r)$ is chosen in Fig.~\ref{CoulSW}, with a radius $b=0.01$ which is small  compared to the Bohr radius, which is $B=2$ in our units.  If alone, this potential $\lambda v(r)$ requires a strength $\lambda_n b^2=(2n-1)^2\pi^2/4$ to support $n$ bound states in S-wave, with numerical values $\{\lambda_n b^2\}=\{2.46, 22.2, \ldots\}$. These are precisely the values at which the atomic spectrum is rearranged in Fig.~\ref{CoulSW}, with the $n$S state falling into the domain of nuclear energies and all other  $i$S atomic states with $i>n$ experiencing a sudden change and drops to (but slightly above) the unperturbed $(i-1)$S energy.

The theory of level shifts of exotic atoms is rather well established, see e.g., \cite[Ch.~6]{Ericson:1988gk}.  The discussion is restricted here to non-relativistic potentials, though exotic atoms have been more recently studied in the framework of effective field theory ~\cite{Holstein:1999nq}. Ordinary perturbation theory is not applicable here. For instance, a hard core of radius $b$ much smaller than the Bohr radius $B$ produces a tiny upward shift of the level, while first-order perturbation theory gives an infinite contribution! The expansion parameter here is \emph{not} the strength of the potential, but the ratio $b/B$ of its range to the Bohr radius, and more precisely, the ratio $a/B$ of its scattering length to the Bohr radius. The scheme of this ``radius perturbation theory'' is outlined in \cite{Mandelzweig:1977cu}. For the sake of this paper, the first order term of this new expansion is sufficient. It is due to Deser et al.~\cite{Deser:1954vq}, Trueman \cite{Trueman61}, etc., and reads
\begin{equation}
\label{eq:Trueman}
{E_n-E_{0,n}\over E_{0,n}}  \simeq-{4\over n}\left({a\over B}\right)~,
\end{equation}
where $a$ is the scattering length in the potential $\lambda v(r)$.  Here, $E_{0,n}$ ($=-1/(4 n^2)$ in our units) is the pure Coulomb energy, and $E_n$ the energy of $n$S level of the modified Coulomb interaction ($n=1, 2,\ldots$). Only in the case where $\lambda v(r)$ is very weak, the scattering length is given by the Born approximation, i.e., $a\propto \lambda$, and ordinary perturbation theory is recovered.  A pedestrian derivation of (\ref{eq:Trueman}) is given in Appendix A. The presence of $a$ instead of $\lambda$ in (\ref{eq:Trueman}) indicates that the strong potential $\lambda v(r)$ acts many times, so that the shift is by no mean a perturbative effect.

The Deser--Trueman formula has sometimes been blamed for being inaccurate. In fact, if the scattering length is calculated with Coulomb interference effects, it is usually extremely good., see, e.g., \cite{Carbonell:1992wd} for a discussion and \cite{Ericson:2003ju} for higher-order corrections. However, this approximation obviously breaks down if the scattering length becomes very large, i.e., if the potential $\lambda v(r)$ approaches the situation of supporting a bound state.

Now the pattern in Fig.~\ref{CoulSW} can be read as follows. For small positive $\lambda$, the additional potential is deeply attractive but produces a small scattering length and hence a small energy shift. As the critical strength $\lambda=\lambda_1$ for binding in $\lambda v(r)$ is approached, the scattering length increases rapidly, and there is a sudden change of the energies. The ground-state of the system, which is an atomic 1S level for small  $\lambda$ and a deeply bound nuclear state for $\lambda \gs \lambda_1$ evolves continuously (from first principles it should be a concave function of $\lambda$, and monotonic if $v(r)<0$ \cite{Thirring:1979b3}).

Beyond the critical region $\lambda\sim \lambda_1$, the scattering length $a$ becomes small again, but \emph{positive}.  Remarkably, the Deser--Trueman formula (\ref{eq:Trueman}) is again valid,
 and accounts for the nearly horizontal plateau experienced by the second state near $E_{0,1}=-1/4$.  A spectroscopic study near $\lambda\gs \lambda_1$ would reveal a sequence of seemingly 1S, 2S, 3S, etc., states slightly shifted \emph{upwards}  though the Coulomb potential is modified by an \emph{attractive} term. This is intimately connected with very low energy scattering: a negative phase-shift $\delta$ can be observed with an attractive potential which has a weakly-bound state, and mimics the effect of a repulsive potential. (The difference will manifest itself if energy increases: the phase-shift produced by a repulsive potential will evolve as $\delta(T)\to 0$ as the scattering energy $T$ increases, while for the attractive potential with a bound state, according to the Levinson theorem, $\delta(T)\to -\pi$.)

The occurrence of an atomic level near $E_{0,1}=-1/4$ for  $\lambda\gs \lambda_1$ can also be understood  from the nodal structure. A deeply-bound nuclear state has a short spatial extension, of the order $b$. To ensure  orthogonality with this nuclear  state,  the first atomic state should develop an oscillation at short distance, with a zero at $r_0\sim b/2$. This zero is nearly equivalent to the effect of a hard core of radius $r_0$. Hence, if $u(r)$ denotes the reduced radial wave function, the upper part of the spectrum evolves from  the boundary 
condition $u(0)=0$ to $u(r_0)=0$, a very small change if $r_0\ll B$.

As pointed out, e.g.,  in Refs.\ \cite{kolomeisky:022721,PhysRevC.26.2381}, the $\delta E_n\propto n^{-3}$ behaviour is equivalent to a constant ``quantum defect''.  For instance, the spectrum of peripheral S-waves excitations of Rydberg atoms is usually written as
\begin{equation}\label{Rydberg}
E_n=-{\hbar^2\over 2 \mu B^2}\,{1\over (n-\nu)^2}~,
\end{equation}
where $B$ is the Bohr radius, $\mu$ the reduced mass, and $\nu$ defines the quantum defect. A constant $\nu$ is equivalent to $\delta E_n\propto n^{-3}$, as for the Trueman formula (\ref{eq:Trueman}).
Indeed, if the excitation of the inner electron core is neglected, the dynamics is dominated by the Coulomb potential $-1/r$ felt by the last electron, which becomes stronger than $-1/r$ when this electron penetrates the core. Within this model, one can vary the strength of this additional attraction from zero to its actual value, or even higher, and it has been claimed that the Zel'dovich effect can be  observed in this way, especially at high $n$ \cite{kolomeisky:022721}.

\section{The limit of a point interaction}\label{se:pt-int}
The simplest solvable model of exotic atoms is realised with a zero-range interaction. The formalism of the so-called ``point-interaction'' is well documented, see, e.g., \cite{alb}, where the case of a point-interaction supplementing the Coulomb potential is also treated, without, however, a detailed discussion of the resulting spectrum.

It is known that an attractive delta function leads to a collapse in the Schr\"odinger equation. In more rigorous terms, the Hamiltonian should be redefined to be self-adjoint. For S-wave, a point interaction  of strength $g=1/a$, located at $r=0$, changes the usual boundary conditions $u(0)=0$, $u'(0)=1$ (possibly modified by the normalisation) by $u'/u=1/a$ at $r=0$. Note that $a$ is the Coulomb-corrected scattering length.

\begin{figure}[!htpc]
\begin{minipage}{.65\textwidth}
\centering
\scalebox{.6}{\includegraphics{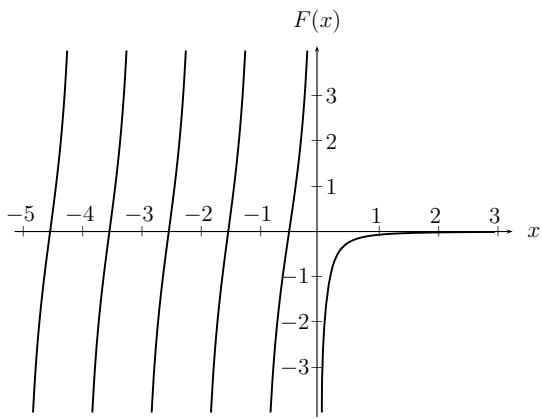}}
\end{minipage}
\begin{minipage}{.34\textwidth}
\caption{\label{FunctionF} Graph of the function $F$ used to calculate the 
spectrum for a Coulomb potential supplemented by a point interaction.}
\end{minipage}
\end{figure}

\begin{figure}[!htpc]
\begin{minipage}{.7\textwidth}
\centering
\scalebox{.60}{\includegraphics{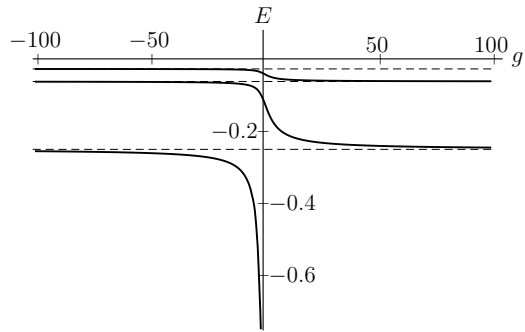}}
\end{minipage}
\begin{minipage}{.28\textwidth}
\caption{\label{Fig2} First few energy levels in a Coulomb potential 
modified by a point interaction  of strength $g$. The dotted 
lines correspond to the pure Coulomb levels $E_{0,n}$.}
\end{minipage}
\end{figure}

In this model, the S-wave eigenenergies are given by $u'(0)/u(0)=1/a$ applied to the reduced radial wave function of the pure Coulomb problem,
which results into \cite{alb}
\begin{equation}\label{eq:basiceq}
F(-2/k)=1/a~,\quad F(x)=\Psi(1+x)-{1\over2}\ln(x^{2})-{1\over 2x}~,
\end{equation}
in terms of the digamma function $\Psi(x)=\Gamma'(x)/\Gamma(x)$. 
Using the reflection formula \cite{Abramovitz64}, the function $F$ can be rewritten as
\begin{equation}
F(-x)=\pi\cot(\pi x)+\Psi(x)-{1\over2}\ln(x^{2})+{1\over 2x}~,
\end{equation}
explaining the behaviour observed on the left-hand side of
Fig.~\ref{FunctionF}.
Equation (\ref{eq:basiceq}) shows that $a \to -\infty$ corresponds to the plain 
Coulomb interaction, where $E_{n}=E_{0,n}$. 
For small deviations, the Trueman formula (\ref{eq:Trueman}) can  be 
recovered form Eq.~(\ref{eq:basiceq}), as shown in \cite{alb}.
The behaviour of the first $n$S levels is displayed in Fig.~\ref{Fig2},  for
$a$ increasing from this limit: a sharp changes is clearly seen
near $1/a =0$, beautifully illustrating the Zel'dovich effect.

A comprehensive analytic treatment of the Zel'dovich effect has been given by Kok et al.\ \cite{PhysRevC.26.2381} using a delta-shell interaction $v(r)\propto -\delta(r-R)$, both for S-waves and higher waves ($\ell>0$).
\section{Rearrangement with square wells}\label{se:simplemod}
\subsection{Model}
The  patterns of energy shifts experienced by exotic atoms when the strength of nuclear potential increases can be studied in a simplified model where the three-dimensional Coulomb interaction is replaced by a one-dimensional square well supplemented by a narrow square well in the middle: the odd-state sector has the same type of rearrangement as the exotic atoms, while the even sector shows a new type of rearrangement. The effect of symmetry breaking  can be  studied by moving the attractive spike aside from the middle.

\begin{figure}[!ht]
\begin{minipage}{.65\textwidth}
\centerline{\scalebox{.8}{\includegraphics{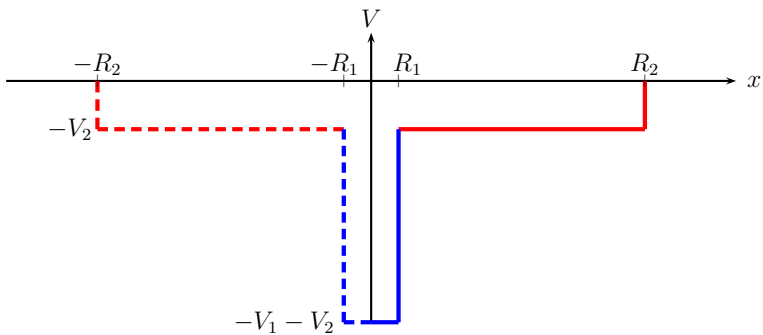}}}
\end{minipage}
\begin{minipage}{.34\textwidth}
\caption{\label{DW1D} One-dimensional  double square-well}
\end{minipage}
\end{figure}

The potential, shown in Fig.~\ref{DW1D}, reads
\begin{equation}\label{double-well}
V(x)=-V_1 \theta(R_1^2-x^2) -V_2\, \theta(R_2^2-x^2)~,
\end{equation}
with value $-V_1-V_2$ for $0<|x|<R_1$, and $-V_2$ for $R_1<|x|<R_2$ and 0 for $|x|>R_2$, see Fig.~\ref{DW1D}. Slightly simpler would be the case of an infinite square well in which an additional well is digged: it can be proposed as an exercise.

The starting point $V_1=0$ with the model (\ref{double-well}) is an one-dimensional square well of depth $V_2$ and radius $R_2$. Its intrinsic spectral properties depends only on the product $R_2^2V_2$. With a value 80, which is realised in the following examples with $R_2=1$ and $V_2=80$, there are six bound states, three even levels and three odd ones. See, e.g., \cite{bonfim:43} for solving the square well problem.
\subsection{Odd states in a symmetric double well}\label{sub:oddsq}
Besides a normalisation factor $\sqrt2$, the odd sector is equivalent to the S-wave sector in a central potential $V(r)$.
The radial wave function $u(r)$ is thus $u(r)=u_1(r)=\sin(r\sqrt{V_1-k^2})$ for $r<R_1$, and $u(r)=u_2(r)=u_1(R_1) \cos[k'(r-R_1)]+ u'_1(R_1) \sin[k' (r-R_1)]/k'$ if $R_1<r<R_2$ with $k'^2=V_2-k^2$, and suitable changes
$\sin\to\sinh$ and $\cos\to\cosh$ if $k^2> V_2$. The eigenenergies can be obtained by matching this intermediate solution $u_2$ to  the external solution $u_3(r)=\exp(-k r)$ at $r=R_2$  , i.e., imposing $u_2(R_2)u'_3(R_2)-u'_2(R_2)u_3(R_2)=0$.  The calculation involves only elementary trigonometric functions, and the spectrum can be computed easily.

\begin{figure}[!ht]
\centerline{\scalebox{.7}{\includegraphics{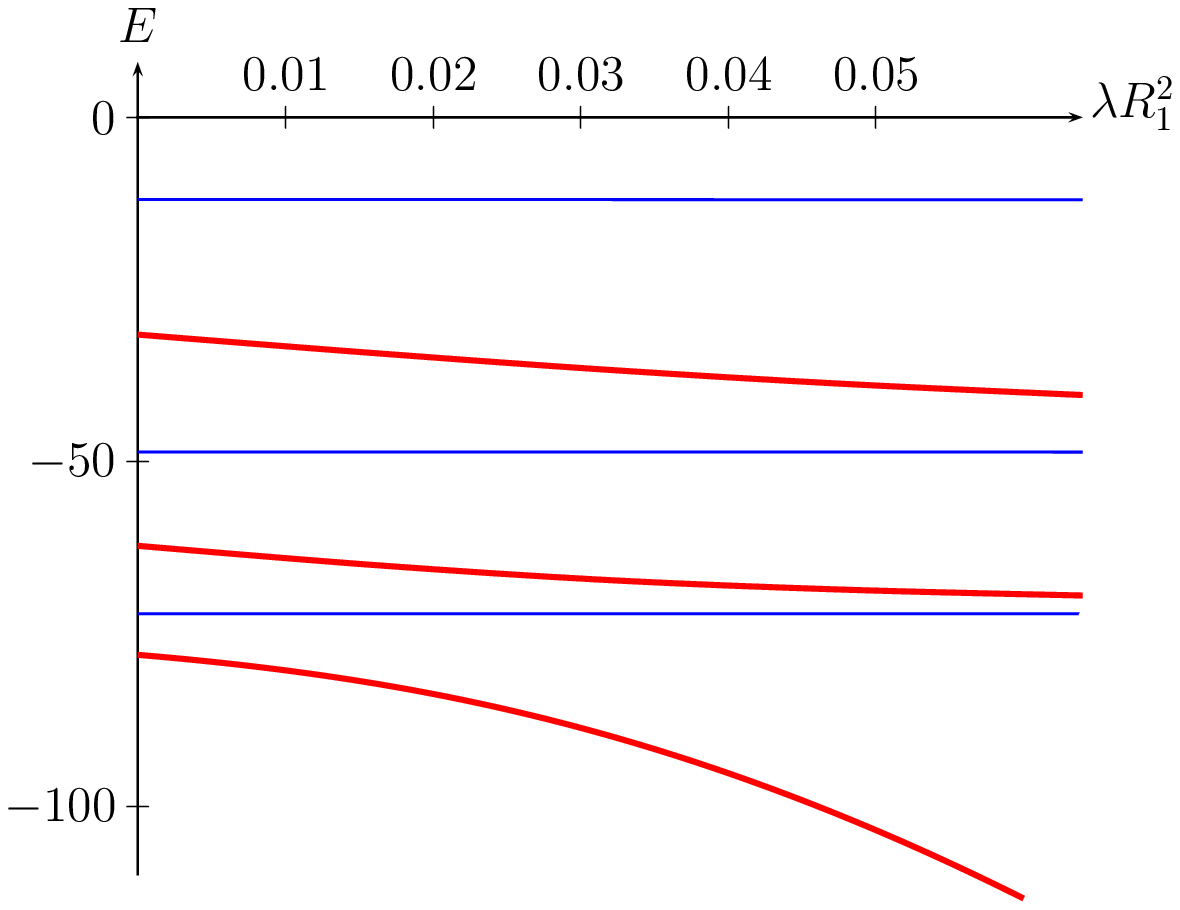}}\hspace*{.3cm}\scalebox{.7}{\includegraphics{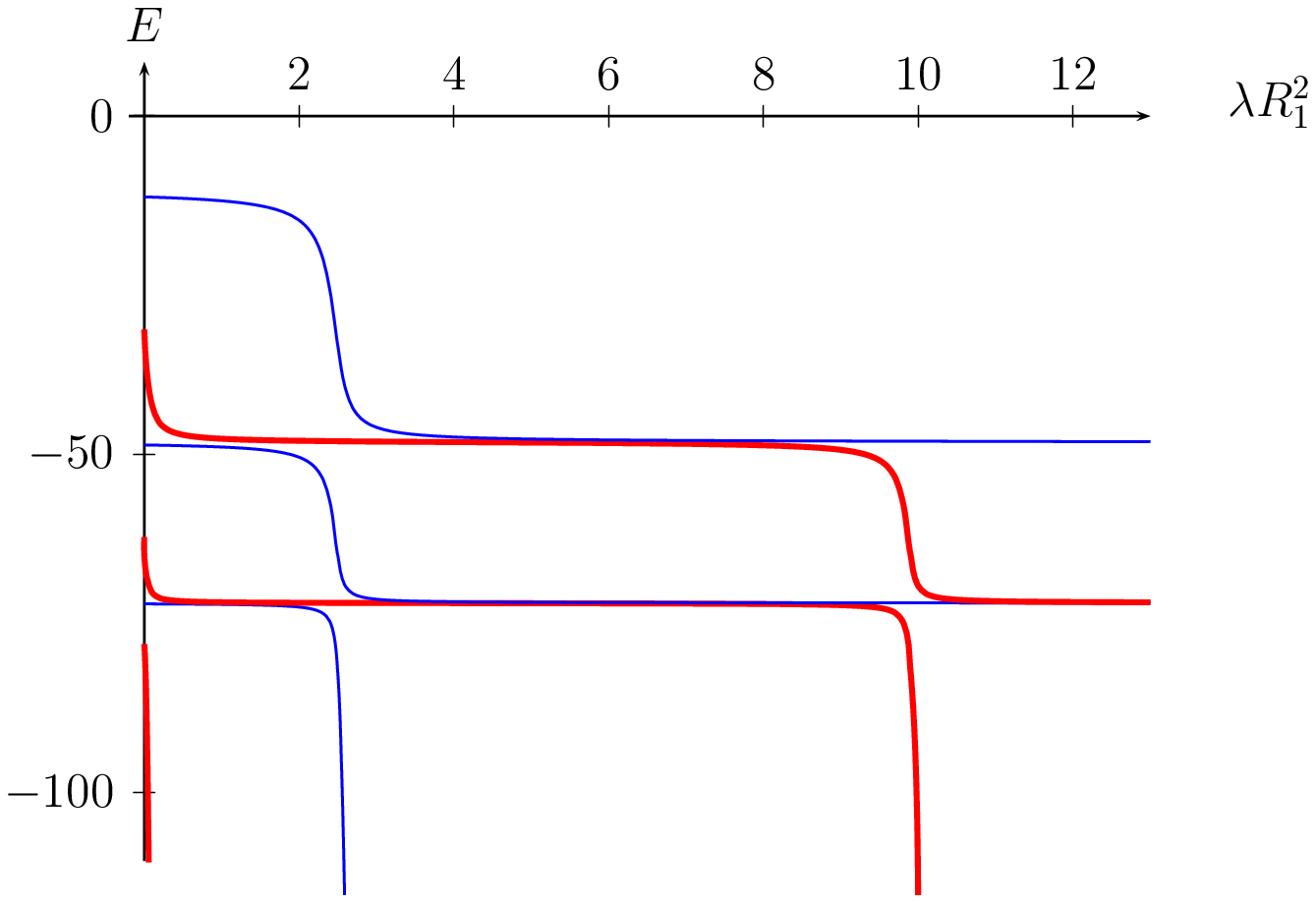}}}
\caption{\label{OddEvenSq} Level rearrangement of the odd (thin line) and even (thick line) states of the double square-well, with $R_2=1$, $V_2=80$, $R_1=0.01$ and increasing $V_1$.}
\end{figure}

The energy levels as functions of $V_1$ are displayed in Fig.~\ref{OddEvenSq}. The rearrangement pattern is clearly seen, and is especially pronounced if $R_1\ll R_2$. The difference from the Coulomb case is that, for the square well,  when a bound state collapses from the ``atomic'' to the ``nuclear'' energy range, a new state is created from the continuum.
\subsection{Even states in a double well}\label{sub:evensq}
The even spectrum of the potential (\ref{double-well}) is  given by 
$w(x)=w_1(x)=\cos(x\sqrt{V_1-k^2})$ for $0\le x<R_1$, and $w(x)=w_2(x)=w_1(R_1) \cos[k'(x-R_1)]+ w'_1(R_1) \sin[k' (x-R_1)]/k'$ if $R_1<x<R_2$ with $k'^2=V_2-k^2$, and suitable changes
$\sin\to\sinh$ and $\cos\to\cosh$ if $k^2> V_2$. Then the matching to $w(x)=w_3(x)=\exp(- k x)$ gives the eigenenergies.

The results are shown in Fig.~\ref{OddEvenSq}, with the same parameters as for the odd part. The same pattern of ``plateaux'' is seen as for the odd parts, with, however, some noticeable differences: 
\begin{itemize}
\item
In quantum mechanics with space dimension $d=1$ (actually for any $d\le2$), any attractive potential supports at least one bound state. In particular, a nuclear state develops in the
 narrow potential of width $2R_1$  even for arbitrarily small values of its depth $V_1$. Hence the ground-state level starts immediately falling down as $V_1$ increases from zero,
\item
The first even excitation does not stabilise  near the value of the unperturbed even ground state, it reaches a plateau corresponding to the first unperturbed \emph{odd} state.
\item
Similarly, each higher even level acquires an energy corresponding to the neighbouring  unperturbed odd level.
\item
When $V_1R_1^2$ reaches about 2.46, enabling the narrow square well to support a second state, a  new rearrangement is observed, with, again,  values close to these of the unperturbed odd spectrum.
\end{itemize}

In short, the energies corresponding to the even states of the initial spectrum quickly disappear. The energies corresponding to the odd states remain, and become almost degenerate, except when a rearrangement occurs.

\begin{figure}[!hc]
\begin{minipage}{.65\textwidth}
\centerline{\scalebox{.65}{\includegraphics{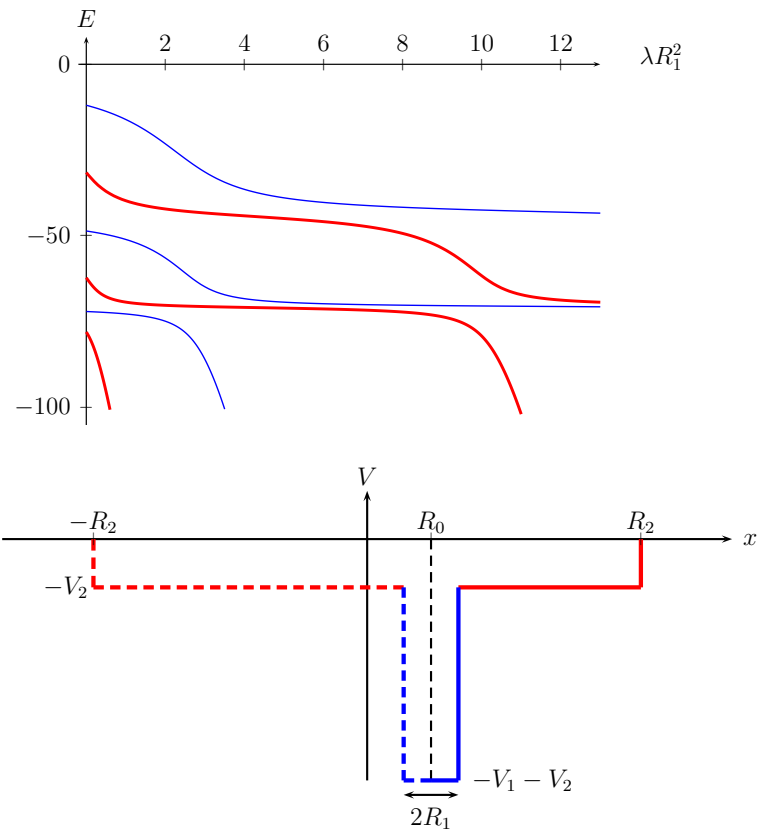}}}
\end{minipage}
\begin{minipage}{.34\textwidth}
\caption{\label{OddEvenSqa} Same as Fig.~\protect\ref{OddEvenSq}, but for a wider range $R_1=0.1$ for the additional potential.}
\end{minipage}
\end{figure}

The degeneracy observed in Fig.~\ref{OddEvenSq} depends crucially on the addtional potential being of very short range. 
For comparison the case of a wider range $R_1=0.1$ is shown in Fig.~\ref{OddEvenSqa}. Though the rearrangement pattern is clearly visible, the transition is much smoother, and the  almost degeneracy limited to smaller intervals of the coupling constant $V_1$, and less pronounced. 

\subsection{Spectrum in an asymmetric potential}
\begin{figure}[!ht]
\begin{minipage}{.65\textwidth}
\centerline{\scalebox{.8}{\includegraphics{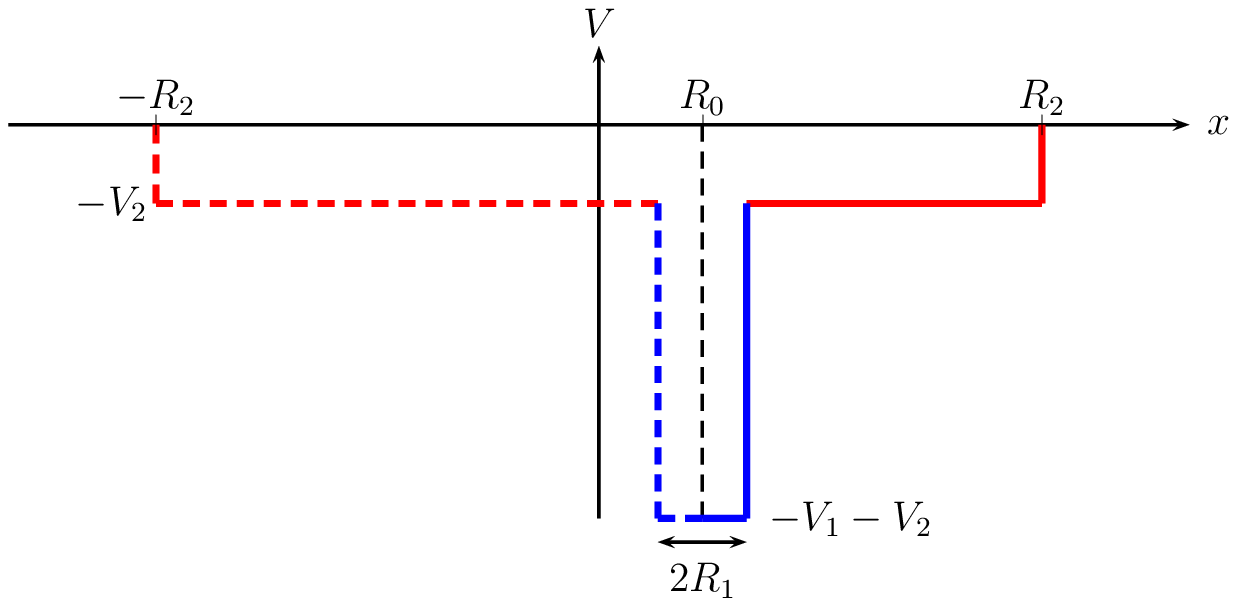}}}
\end{minipage}
\begin{minipage}{.34\textwidth}
\caption{\label{DW1DaP} Asymmetric double square-well}
\end{minipage}
\end{figure}

\begin{figure}[t]
\begin{center}
\centerline{\scalebox{.7}{\includegraphics{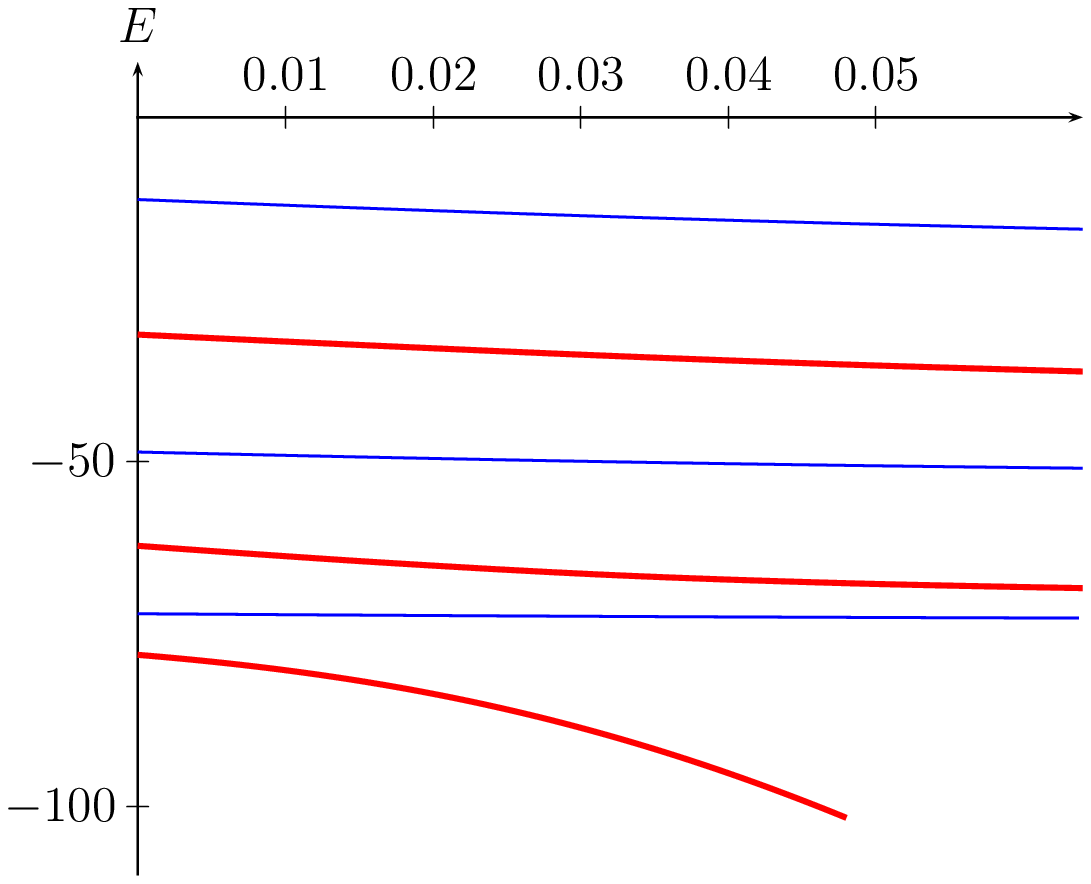}}\hspace*{.3cm}\scalebox{.7}{\includegraphics{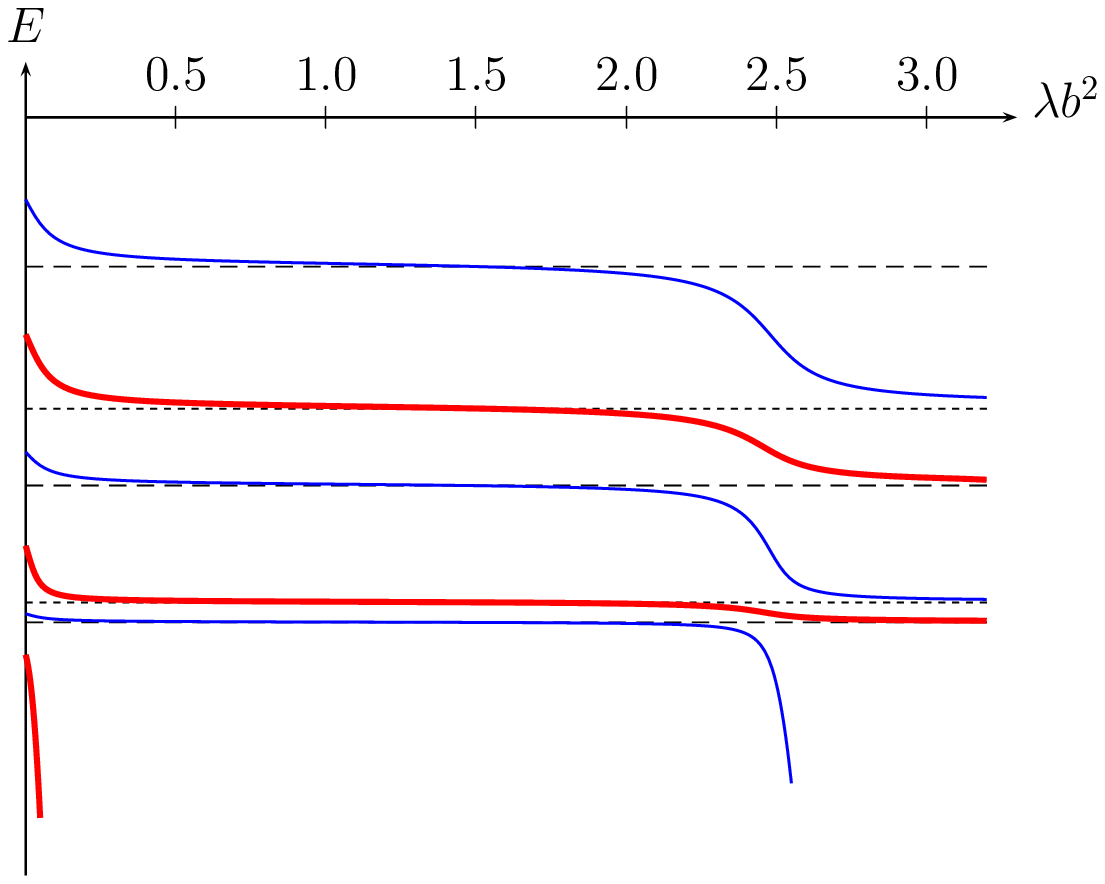}}}
\end{center}
\caption{\label{DW1DaSp} Spectrum in the asymmetric double square-well of Fig.\ \protect\ref{DW1DaP}, with $R_2=1$, $V_2=50$, $R_1=0.01$, $R_0=0.1$ and variable $V_1$.
The dotted lines correspond to the states in the box $[-R_2,R_0]$, and the dashed ones to those in the box $[R_0,R_2]$, all with depth $V_2$ and hard wall at $R_0$.}
\end{figure}

To check the interpretation of the patterns observed for the odd and even parts of the spectrum,  let us break parity and consider the asymmetric double well of Fig.~\ref{DW1DaP}.
For the sake of illustration, the centre of the spike is taken at $R_0=0.1$. The spectrum, as a function of $V_1$, is displayed in Fig.~\ref{DW1DaSp}:
Plateaux are observed, again, with energy values corresponding approximately to the combination of $(i)$  the spectrum in a well of depth $-V_2$ between $x=R_0+R_1$ and $x=R_2$ and a hard core on the left, i.e., a boundary condition $w(R_0+R_1)=0$, and  $(ii)$  the spectrum in a well of depth $-V_2$ between $x=-R_2$ and $x=R_0-R_1$  with $w(R_0-R_1)=0$.
%

\begin{figure}[t]
\centerline{\scalebox{.7}{\includegraphics{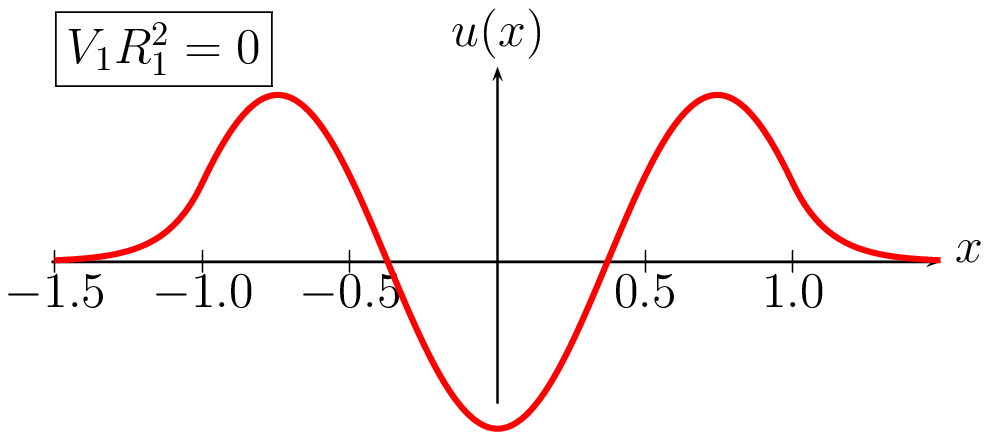}}\hspace*{.3cm}\scalebox{.7}{\includegraphics{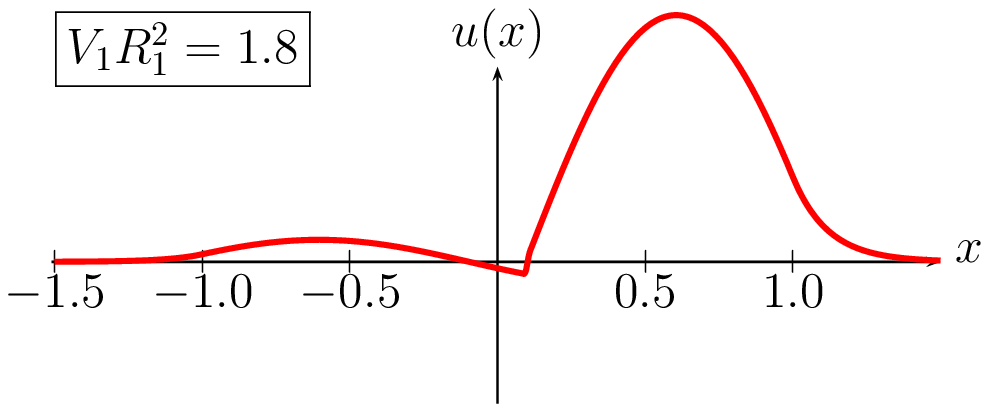}}}
\centerline{\scalebox{.7}{\includegraphics{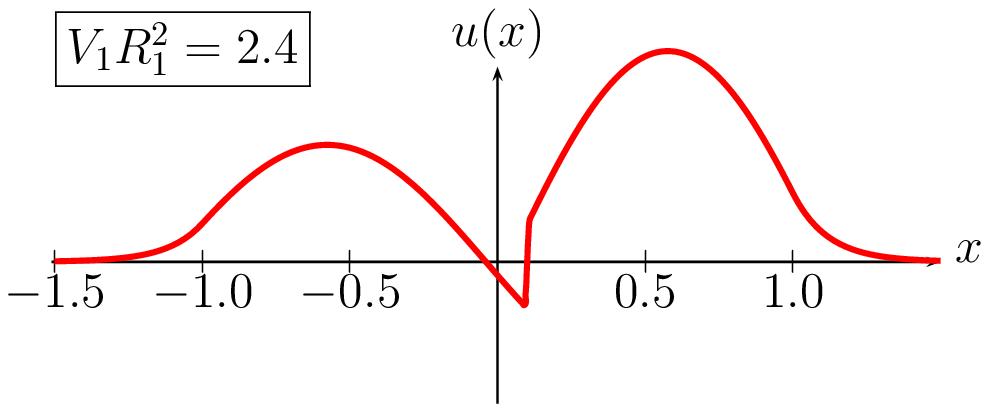}}\hspace*{.3cm}\scalebox{.7}{\includegraphics{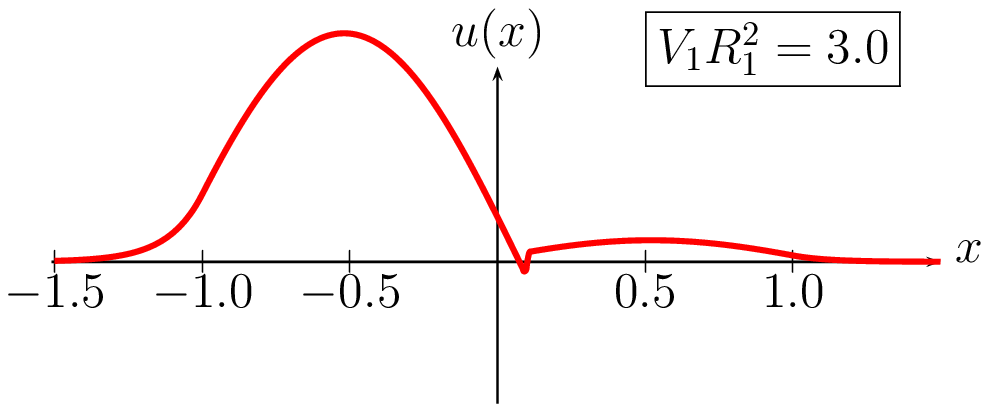}}}
\caption{\label{DW1DaWf} Wave function of the third level in the asymmetric double square-well of Fig.\ \protect\ref{DW1DaP}, with $R_2=1$, $V_2=80$, $R_1=0.01$, $R_0=0.1$ and variable $V_1$ near a rearrangement. An enlargement of the region near $x=0.1$ would confirm that the wave function and its derivative are continuous.}
\end{figure}

It is interesting to follow how the wave function evolves when a rearrangement occurs. In Fig.\ \ref{DW1DaWf}, the third level is chosen. For $V_1=0$, it is the first even excitation with energy $E_3\simeq -62.18$, and the wave function $u(x)$ is the usual sinus function matching exponential tails. 
On the first plateau, with energy near $-70$, this wave function  is almost entirely located on the right. 
As rearrangement takes place, the probability is shared by both sides.  On the second plateau, with a energy near $-73$ corresponding to the ground state in the wider part with hard wall at $R_0$, the 
wave function is mostly on the left.

When the narrow well has only deeply bound states,  it acts as  an effective hard wall between the two boxes, at the right and and the left of $R_0$. However, when a new state occurs with a small energy and  an extended wave function, it opens the gate, and states can move from the right to the left, or vice-versa.

It is possible to study how the spectrum in Fig.~\ref{DW1DaSp}  evolves if the centre of the spike moves to the right, i.e., $R_0\to R_2-R_1$: the dotted line move up and disappear, while the dashed lines move down and become more numerous.  Eventually, if the depth $V_1$ is large, the spectrum becomes very similar to the odd part of the spectrum in Figs.~\ref{DW1D}, \ref{OddEvenSq}, except for a change $R_1\to 2 R_1$ and $R_2\to 2 R_2$.; This illustrates again that for the upper part of the spectrum, a deep hole is equivalent to a hard wall.

\section{Rearrangement in quantum dots}
\subsection{Level rearrangement in an harmonic well}
There is a considerable recent literature on quantum dots \cite{0034-4885-64-6-201}, usually dealing with many particles in a trap, with a magnetic field. Let us consider the simplified problem of two particles confined by a wide harmonic trap, and interacting with short-range forces, 
\begin{equation}\label{eq:dot1}
H={\vec{p}_1^2\over 2m}+{\vec{p}_2^2\over 2 m}+ K r_1^2+Kr_2^2+\lambda v(|\vec{r}_2-\vec{r}_1|)~.
\end{equation}
The centre-of-mass oscillates in a pure harmonic potential, and the separation $\vec{r}=\vec{r}_2-\vec{r}_1$ is governed by 
\begin{equation}\label{eq:dot2}
h={\vec{p}^2\over m}+ K r^2+\lambda v(r)~,
\end{equation}
If $v(r)$ is attractive or, at least, has attractive parts, $\lambda v(r)$ will support bound states for  large enough $\lambda$. 
The same phenomenon of level rearrangement is observed,  as shown in the simple example of harmonic oscillator and square well.  As for the case of exotic atoms, the effect of  ``level repulsion'' is observed, that ovoids any crossing of trajectories corresponding to the same orbital momentum.
\begin{figure}[!htpc]
\begin{minipage}{.65\textwidth}
\begin{center}\scalebox{.75}{\includegraphics{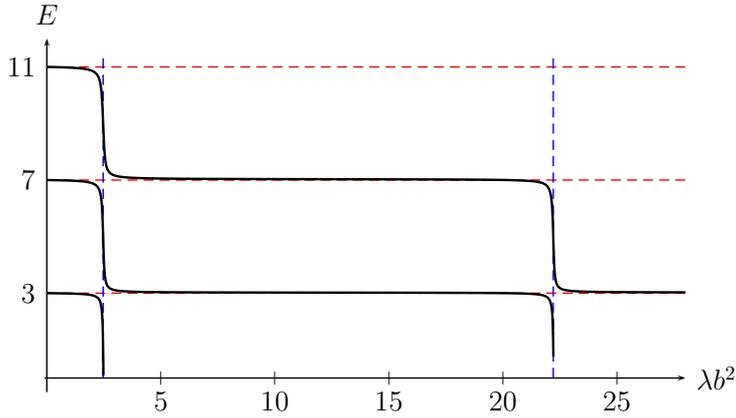}}
\end{center}
\end{minipage}
\begin{minipage}{.33\textwidth}
\caption{\label{HOrearr} First  few S-wave energy levels for an harmonic oscillator supplemented by a square well of variable strength, i.e., $V(r)=r^2-\lambda \Theta(b-r)$ and radius $b=0.01$. The vertical lines correspond to the coupling $\lambda_1=\pi^2/4$ and $\lambda_2=9 \lambda_1$ at which a first and a second S-wave bound state occurs in the square well alone.}
 \end{minipage}
\end{figure}
\subsection{Dependence upon the radial number}
As for the theory, it is similar to that of exotic atoms. The analogue of the Trueman--Deser formula, for any long-range potential combined with a short-rangfe potential, reads
\begin{equation}\label{eq:dot3}
\delta E_n \simeq 4\pi a\,|\phi_n(0)|^2 ~,
\end{equation}
indicating that the energy shift is proportional to the square of the value at the origin of the wave function of the pure long-range potential.
It is worth pointing that the dependence upon the radial number $n$ is different for the Coulomb and the oscillator problems:
\begin{itemize}
\item
For a narrow pocket of attraction added to an harmonic confinement, the energy shifts at large $n$ increase as $n^{1/2}$, since the square of the wave function at the origin is $|\phi_n(0)|^2=8/[\sqrt{\pi}B(n+1,1/2)]$, where $B$ is the beta function. But for very arge enough $n$, the first nodes of the radial function come in the range of $v(r)$, and then $\delta E$ decreases with $n$. Moreover,  for very large $n$, the radial Schr{\"o}dinger equation is dominated at short distance by the energy term.
\item 
For a Coulomb interaction,  $|\phi_n(0)|^2\propto n^{-3}$, and hence $\delta E_n\propto n^{-3}$, a well-known property of exotic atoms. As explained, e.g., in a  review article on protonium \cite{Klempt:2002ap} and briefly explained in Appendix, the first node of the $n$S radial function, as $n$ increases, does not go to $0$. In the case $\hbar^2/(2\mu)=e^2=1$, the node of the 2S level is at $r=4$, while the first node of $n$S at large $n$ is at  $r\simeq 3.67$. Hence the Coulomb wave function never exhibits nodes within the range of the nuclear potential. Moreover, the energy term  is always negligible in comparison with $\lambda v(r)$ at short distances.
\end{itemize}

\subsection{From Coulomb to harmonic rearrangement}
The KS transformation \cite{Mavromatis:1998} relates Coulomb and harmonic-oscillator potentials.
The radial equation for a Coulomb system in three dimension (with  $\hbar=2\mu=1$) reads
\begin{equation}\label{eq:KS1}
-u''(r)+{\ell(\ell+1)\over r^2} u(r)-{\alpha\over r} u(r) - E u(r)=0~,
\end{equation}
with $u(0)=0$ and $u(r)\to 0$ as $r\to \infty$ becomes
\begin{equation}\label{eq:KS2}
-\phi''(\rho)+{L(L+1)\over \rho^2}\phi(\rho) + 4 (-E) \rho^2 \phi(\rho) - 4 \alpha \phi(\rho)=0~,
\end{equation}
if  $r=\rho^2$, $u(r)=\rho^{1/2}\phi(\rho)$, and $L=2\ell+1/2$. The modified angular momentum
 can be interpreted as  relevant in a higher-dimensional world  \cite{Mavromatis:1998}. But Eq.~(\ref{eq:KS2})  is precisely the Schr{\" o}dinger equation for the three-dimensional oscillator with (fixed) energy $4\alpha$ and oscillator strength $4(-E)$ (which is positive), i.e.,
\begin{equation}\label{eq:KS4}
4\alpha =\sqrt{-4E}(3+4n+2 L)~,\quad n=0,\,1, \ldots~,
\end{equation}
which is equivalent to the Bohr formula
\begin{equation}\label{eq:KS5}
E=-{\alpha^2\over 4(1+n+\ell)^2}~,
\end{equation}
where $1+n+\ell$ is the usual principal quantum number of atomic physics.

Now, an additional potential $\lambda v(r)$ in the Coulomb equation results into a short-range term 
$4 \lambda \rho^2v(\rho^2)$ added to the harmonic oscillator, and all results obtained for exotic atoms translate into the properties listed for a narrow hole added to an harmonic well.

Note that the $n$ dependence is also explained. In the KS transformation, the energy $E$ ($E<0$) in the Coulomb system becomes the strength $-4E$ of the oscillator, while four times the fine structure constant, i.e.,  $4\alpha$ ($\alpha>0$ for attraction) becomes the energy  eigenvalue of the oscillator with angular momentum $L$. If $n$ increases, the oscillator deduced from the KS transformation becomes looser, and hence less sensitive to the short range attraction $4\lambda\rho^2v(\rho^2)$. To maintain a fixed oscillator strength, one should imagine a different Coulomb system for each $n$, with $\alpha\propto n$, hence a Bohr radius independent of $n$, and a wave function at the origin $|\phi(0)|^2\propto n^{-1}$ instead of $n^{-3}$ in the usual case. Then, in this situation,  $\delta E\propto n^{-1}$ for the Coulomb system, and $\delta \epsilon\propto n^{1/2}$ for the harmonic oscillator.
\section{Rearrangement and level ordering}\label{se:order}
In the above examples, there is an interesting superposition of potentials with different level-ordering properties. A square well potential, if deep enough to support many bound states, has the ordering \cite{Landau}
\begin{equation}\label{eq:sqorder}
1S<2P<3D<2S<\ldots~.
\end{equation}
We are adopting here the same notation as in atomic physics is adopted, i.e.,  2P is the first P-state, 3D the first D-state, etc. The Coulomb potential, on the other hand, exhibits the well-known degeneracy
\begin{equation}\label{eq:cborder}
1S<2S=2P<3S=3P=3D<\ldots~,
\end{equation}
while for the harmonic-oscillator case,
\begin{equation}\label{eq:hoorder}
1S<2P<2S=3D<\ldots~,
\end{equation}
with equal spacing.

\begin{figure}[!htbc]
\begin{minipage}{.65\textwidth}
\centerline{\scalebox{.6}{\includegraphics{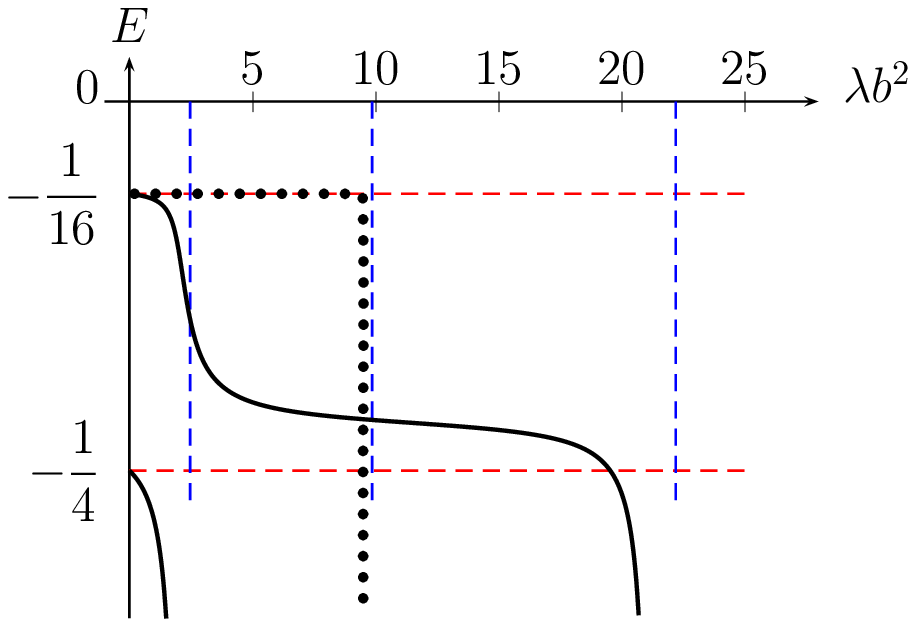}}
\hspace*{-.1cm}
\scalebox{.6}{\includegraphics{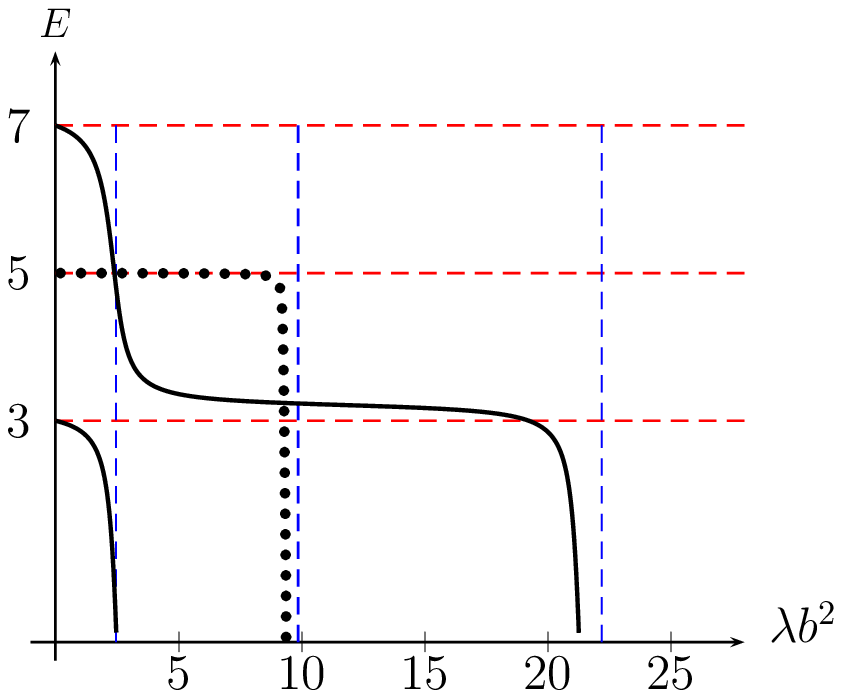}}\hfill}
\end{minipage}
\begin{minipage}{.34\textwidth}
\caption{\label{fig:crossing}1S and 2S levels (solid line) and 2P level (dotted line), for Coulomb (left) or harmonic-oscillator (right) potential plus a square well of radius $b=0.1$ of increasing strength $\lambda$. The horizontal lines are the unperturbed values, the vertical ones indicate the strength $\lambda$ at which the square well alone starts supporting a new bound state.}
\end{minipage}
\end{figure}

The pattern of 1S, 2S and 2P levels for Coulomb (left) or harmonic oscillator (right) supplemented by a short-range square well of increasing strength is given in Fig.~\ref{fig:crossing}.

In the Coulomb case, the degeneracy is broken at small $\lambda$ as $E(2S)<E(2P)$ since the 2P wave function vanishes at $r=0$. The 2S drops when the 1S state falls into the region of deep binding. However, the 2P state becomes bound into the square well near $\lambda b^2=\pi^2$, earlier than the 2S for which this occurs near $\lambda b^2=9 \pi^2/4$. This explains the observed crossing.

In the harmonic oscillator case, there is a remarkable double crossing. The 2S drops by the phenomenon of rearrangement, and crosses the 2P level which is first almost unchanged. When the 2P level becomes  bound by the square-well, it crosses again the 2S, which falls down for higher strength.

Note that those patterns  do not contradict the general theorems on level ordering, which have been elaborated in particular for understanding the quarkonium spectra in potential models \cite{Quigg:1979vr,Grosse:1997xu}. If the square well $\lambda v$ is considered as the large $n$ limit of $\lambda v_n(r)=-\lambda/ [1+ (r/b)^n]$, the Laplacian $\Delta v_n=(r v_n)''/r$ can be calculated explicitly, and is easily seen to be positive for small $r$ and negative for large $r$. Hence the theorem  \cite{Quigg:1979vr,Grosse:1997xu} stating that $E(2P)<E(2S)$ if $\Delta V>0$ and  vice-versa cannot be applied here. In our case $V=-1/r+\lambda v$, with the Coulomb part having a vanishing Laplacian, or $V=r^2+\lambda v$, with $\Delta(r^2)>0$.

Figure \ref{fig:crossing} clearly indicates that the rearrangement is much sharper for P-states that for 
S-states. The study could be pursued  for higher value of the orbital momentum and the rearrangement would be observed to become even shaper.

\section{Outlook}
In this article, some remarkable spectral properties of the Schr\"odinger equation have been exhibited, which occur when a strong short-range interaction is added to a wide attractive well. When the short-range part is deep enough to support one or more bound states, it acts as  repulsive barrier on the upper part of the spectrum. Thus the low-lying levels are approximately those which are in  wide well, with, however, the condition that the wave function vanishes in the region of strong attraction. 

It is interesting to follow the spectrum as a function of the strength of the additional short-range attraction. The energy curve exhibit sharp transitions from intervals where they vary slowly. This is the phenomenon of  level-rearrangement, discovered years ago, and generalised here.

It is worth pointing out an important difference between one and higher dimensions regarding rearrangement phenomenon. Since in one dimension, one has the inequality $E_{n-1}<E_n$,  there cannot be any crossing of levels during rearrangement. However, while in
higher dimensions, there cannot be any crossing between levels with same angular momentum, several crossings of levels with different angular momentum will normally occur.

Most applications in the literature deal with exotic atoms, but the phenomenon was first revealed in the context of condense-matter physics, and could well find new applications there. Layers could  be combined, with a variety of voltages, and a variety of interlayer distances, and the situation can perhaps be realised where a tiny change of one of the voltage could provoke a sudden change of the bound state spectrum.

The problem of particles in a trap, with individual confinement and an additional pairwise interaction, has stimulated a copious literature, but the level rearrangement occurring at the transition from individual binding to pairwise binding was never underlined, at least to our knowledge. 

Several further investigations could be done. The problem of absorption has already been mentioned, and it is our intent to study it in some detail. The subject is already documented in the case of exotic atoms, as pions, kaons and especially antiprotons have inelastic interaction with the nucleus. It has been shown that the phenomenon of rearrangement disappears  if the absorptive component of the interaction becomes too strong. See, e.g., \cite{Badalyan:1982,Gal:1996pr} and refs.\ there.

It could be also of interest to study how the system behave, as a function of the coupling factors, if two or more attractive holes are envisaged inside a single wide well.

\appendix
\section{Trueman--Deser formula}
We give here a pedestrian derivation of the Trueman--Deser formula.
Consider a repulsive interaction  added to a long-range attractive potential $V_0(r)$ in unit such that $\hbar^2/(2m)=1$. This short-range repulsion, at energy $E\simeq0$ is equivalent to a hard core potential of radius $a$, where $a$ is the scattering length of $V$. Hence the pure Coulomb and the modified Coulomb problems results for orbital momentum $\ell=0$ into
\begin{eqnarray}
&-u_0''(r)+ V_0(r) u_0(r) =E_0 u_0(r)~,\quad & u_0(0)=0~,\quad u_0(\infty)=0~,\nonumber\\
&-u''(r)+V_0(r) u(r)=E u(r)~,\quad &u(a)=0~,
\quad u (\infty)=0~.
\end{eqnarray}
After multiplication by $u$ and $u_0$, respectively, the difference leads to
\begin{equation}
(E-E_0)\int_a^\infty u_0(r)u(r)\,\mathrm{d}r=u'(a)u_0(a)~.
\end{equation}
In the LHS, the integral is close to the normalisation integral of $u_0$ or $u$, i.e.,  close to unity. If $V_0(r)$ is smooth, then $u(r)$ does not differ much from the shifted version $u_0(r+a)$ of the unperturbed solution. Hence
$u'(a)\simeq u_0(0)$.  Also $u_0$ is nearly linear near $r=0$, and $u_0(a)\simeq u'(0) a$, and eventually
\begin{equation}
E-E_0\simeq u'(0)^2 a~.
\end{equation}
which reduces to (\ref{eq:Trueman}) if $V_0(r)=-1/r$. For a moderately attractive potential, $a$ is negative, but the formula and its derivation remain valid.

For a Coulomb potential, the square of wave function at the origin of the $n$S state, $|\phi_n(0)|^2=u'_n(0)^2$, decreases like $1/n^3$, and so does the energy shift, a property which is well known for exotic atoms.

The $n$-dependence of $|\phi_n(0)|^2$ has been discussed, e.g., in the context of charmonium physics \cite{Quigg:1979vr,Grosse:1997xu}. For power-law potentials $\epsilon(\alpha)r^\alpha$ ($\epsilon$ is the sign function), $|\phi_n(0)|^2$ increases with $n$ if $\alpha<1$, and decreases if $\alpha>1$. If $\alpha=1$, then $|\phi_n(0)|^2$ is independent of $n$ (after normalisation). 
This can be seen from the Schwinger formula \cite{Quigg:1979vr,Grosse:1997xu}
\begin{equation}
u'(0)^2=\int_0^\infty V'(r) u^2(r)\mathrm{d} r~,
\end{equation}
which is also useful for numerical calculations.

For the harmonic oscillator (rescaled to $-u''(r)+ r^2 u(r)=E\, u(r)$ for S waves), it can be shown that
\begin{equation}
u_n'(0)^2 = 1/B(3/2,3/2+n/2)\sim \sqrt{n}~,
\end{equation} 
 in terms of the Euler function $B(x,y)= \Gamma(x) \Gamma(y)/\Gamma(x+y)$.
 
 Note that the question  of a large $n$ limit has a different answer for the Coulomb and oscillator cases.
 In the former case, the $n$-S radial wave function $u_n(r)$ extends outside when $n$ increases, with an asymptotic decrease (in our normalisation) $\exp(-r/(2 n)$. The $2$S state is $u_2(r)\propto r(4-r)\exp(-r/4)$ has its first (an unique) node at $r_1(2)=4$. As $n$ increases, this first node $r_1(n)$ necessarily decreases, as a consequence of the interlacing theorem, however, $\lim_{n\to\infty}\simeq 3.67$, the first node of the Bessel function which satisfies $y''+y/r=0$, $y(0)=0$. Hence if a potential is short-ranged for 1S, it is also short-ranged for all $nS$ states, and also for states with orbital momentum $\ell>0$. On the other hand, for the harmonic oscillator, all $n$S states have about the same size, with the same asymptotic fall-off $\exp(-r^2/2)$. As $n$ increases, the radial equation is approximately $u''+4n u=0$, with first node $r_1(n)\sim \pi/\sqrt{4n}$. Hence an additional potential whose range is short but finite will feel the node structure of states with very high $n$, and the approximation leading to the generalised Deser--Trueman formula (\ref{eq:Trueman}) ceases to be valid. These considerations hold for an harmonic oscillator with fixed strength.

\begin{acknowledgments}
This work was done in the framework of the CEFIPRA Indo--French exchange programs 1501-2 and 3404-D: Rigorous results in quantum information theory, potential scattering and supersymmetric quantum mechanics.  It is a pleasure to thank A.~Martin and T.E.O.~Ericson for very interesting discussions.
\end{acknowledgments}


 \end{document}